# High Density Matter

**J.R.Stone**[1]
*Department of Physics and Astronomy, University of Tennessee*
*Knoxville TN 37996, USA*
*Department of Physics, University of Oxford*
*Oxford OX1 3PU, United Kingdom*
*E-mail:* j.stone@physics.ox.ac.uk

The microscopic composition and properties of matter at super-saturation densities have been the subject of intense investigation for decades. The scarcity of experimental and observational data has lead to the necessary reliance on theoretical models. However, there remains great uncertainty in these models, which, of necessity, have to go beyond the over-simple assumption that high density matter consists only of nucleons and leptons. Heavy strange baryons, mesons and quark matter in different forms and phases have to be included to fulfil basic requirements of fundamental laws of physics. In this review the latest developments in construction of the Equation of State (EoS) of high-density matter at zero and finite temperature assuming different composition of the matter are surveyed. Critical comparison of model EoS with available observational data on neutron stars, including gravitational masses, radii and cooling patterns is presented. The effect of changing rotational frequency on the composition of neutron stars during their lifetime is demonstrated. Compatibility of EoS of high-density, low temperature compact objects and low density, high temperature matter created in heavy-ion collisions is discussed.

[1] Speaker

## 1. Introduction

One of the central questions of current theoretical physics is what constitutes the structure of matter at high densities and temperatures. This research requires not only a joint effort of nuclear, particle and astrophysics theories, but also the use of the most advanced astrophysical observation data and terrestrial experiments to test the theories.

In the generally accepted QCD phase diagram, expressed in terms of temperature T and baryon chemical potential $\mu_B$, two extreme regions can be identified. At low (zero) temperature and high chemical potential (above 1Gev), the best representation of nuclear matter can be found in the cores of cold neutron stars (NS) and, at somewhat higher temperature, in core-collapse supernovae (CCS). The other extreme is at low (zero) chemical potential and extremely high temperature (several hundreds of MeV). This region of the phase diagram corresponds to conditions in the Early Universe (EU). The adjacent region (HI-I) with lower temperatures and somewhat higher chemical potentials can, in principle, be probed in existing terrestrial heavy ion collisions, at LHC, RHIC and GSI. Extension towards even higher chemical potentials and lower temperatures (HI-II) is expected in planned facilities FAIR at GSI and NICA in Dubna, Russia. One of the key questions is the location of the critical point where the hadronic and quark-gluon phases coexist. There is an extensive beam scan campaign at RHIC, performed by the STAR collaboration, in this direction, but no final conclusion has been reached as yet. The position of the critical point in the QCD phase diagram has important implications for the location of the hadronic-quark phase transition in neutron stars and core-collapse supernovae.

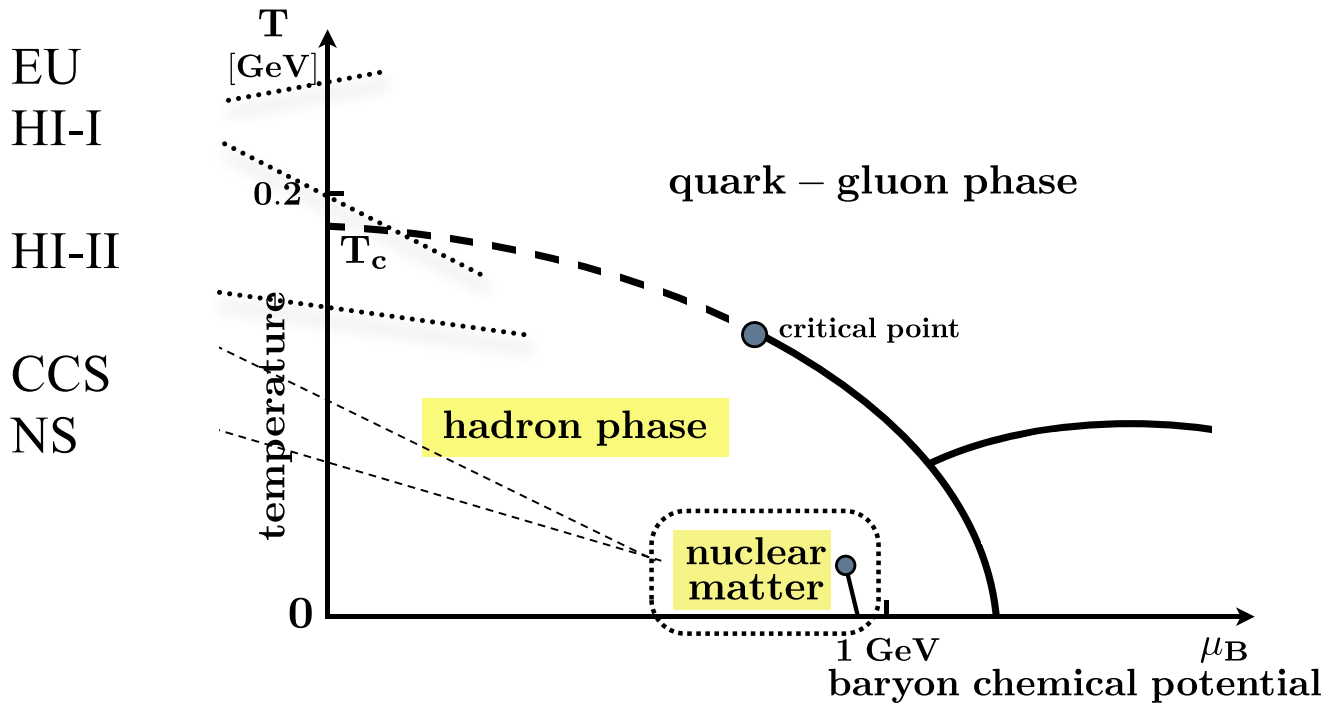

*Figure 1 Simplified QCD phase diagram. Adopted from Ref. [1]. The symbols are explained in text.*

The key property of high density matter is the Equation of State (EoS). Starting from the Boltzmann theory of ideal gases, the relation between the pressure P, energy density ε and temperature T of matter, can be derived. We have

$$P=\rho^2\left(\frac{\partial(\varepsilon/\rho)}{\partial\rho}\right)_{s/\rho}\qquad \varepsilon(\rho,T)=\sum_f\left(\frac{E}{A}(\rho,T)\right)_f\rho_f\qquad \mu_B=(\varepsilon+P)/\rho$$

where $\rho$ is particle number density, *E/A* is the energy per particle and the summation carries over all *f* constituents present in the matter. *E/A* is unknown and has to be determined from nuclear and/or particle physics models. These models are the main source of ambiguity in theoretical determination of the EoS of high density matter. There are many variants of microscopic and phenomenological models of hadronic matter with different levels of complexity. To name a few, mean-field non-relativistic and relativistic models, *ab initio* models with two- and three-body nucleon-nucleon interactions and Quark-Meson coupling models are frequently used [1-5]. There is also a wide choice of composition of hadronic matter, from nucleon only to matter including the full baryon octet and baryon resonances (p, n, Λ, Σ, Ξ, Δ), and mesons (π, K, H-dibaryon condensates). Deconfined (u, d, s) matter has been proposed, which ought to be in a color superconducting state [1,2]. Models of quark matter are even less reliable and range from the different forms of the MIT bag, Nambu-Jona-Lasinio, Chromo-Dielectric, Dyson-Schwinger, perturbative approach to QCD and implementation of the Polyakov-loop technique at non-zero temperature. There are many models of phases of color-superconductivity in quark matter.

The wide freedom in selection and combination of these models in description of the structure of neutron star interiors leads to theoretical prediction of *hybrid* stars, composed of a nucleon-only crust with a core of strange baryons and mesons and stable quark matter in the center or, alternatively, formation of a *strange* star, made of absolutely stable strange quark matter (a configuration of matter more stable than the most stable atomic nucleus, $^{62}Ni$), possibly with a thin nuclear crust of density below the neutron drip threshold. Other suggestions would include *hyperonic* stars. Models of neutron superfluidity and proton superconductivity are also included in the picture.

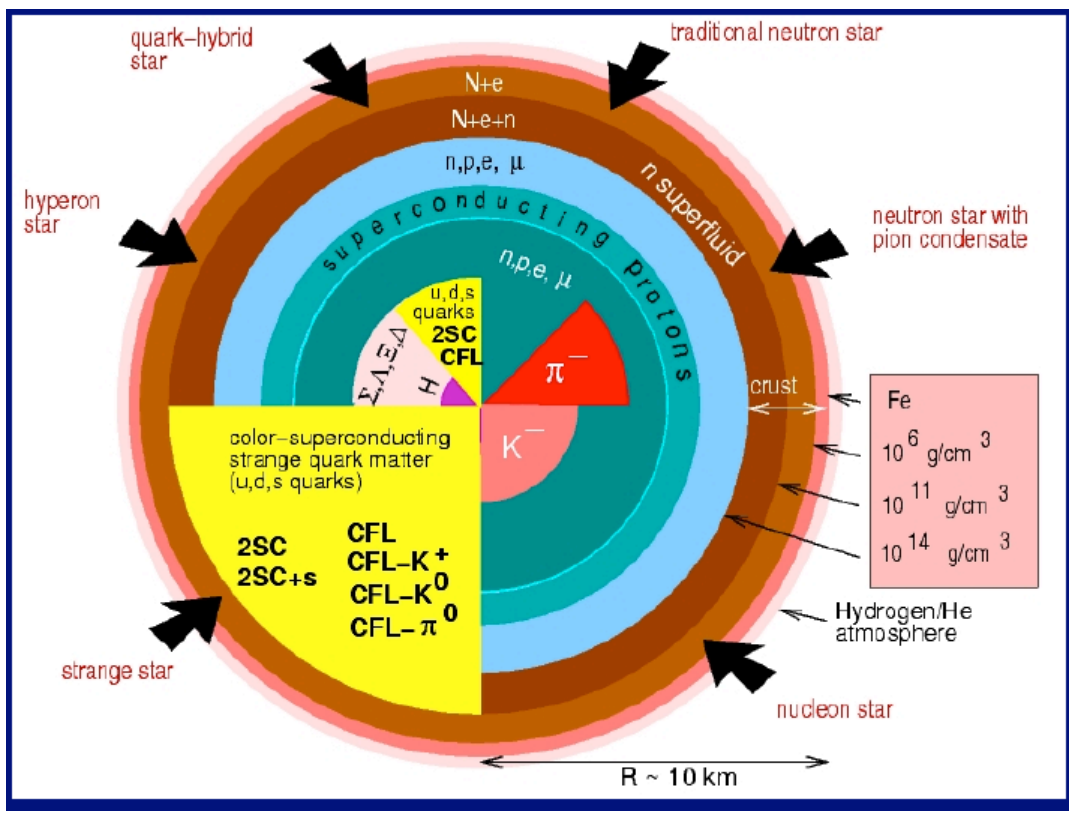

*Figure 2 Complexity of the internal structure of a neutron star. For explanation see Ref. [2].*

All these models, schematically illustrated in Figure 2, are naturally dependent upon a number of adjustable parameters which are constrained by available observational or experimental data. This review concentrates mainly on models of hybrid stars, with critical assessment of the status of available constraints forming the central theme.

## 2. Constraints from astrophysical observations

Selected, firmly established, properties of three objects, (i) neutron stars, (ii) core-collapse supernovae and (iii) proto-neutron stars, which constrain the EoS of high density matter, are considered in detail. More speculative constraints, suggested in connection with the expected observation of gravitational waves, are briefly mentioned.

### 2.1 Neutron stars

Gravitational mass, radius and the period of rotation of neutron stars are the main parameters extracted from observation. The most precise limits on neutron star masses come from radio pulsars in binary systems with another neutron star but binaries with a white dwarf, a main sequence star and a double pulsar binary are also used for mass determination. Neutron stars in X-ray binaries, which accrete material from a companion star and flare to life with a burst of X-rays have become important for determination of not only their mass but also their radius, as discussed below.

#### 2.1.1 Gravitational mass and radii of unrelated objects

Hydrostatic equilibrium of a non-rotating (spherical) neutron star is well described by the general relativistic Tolman-Oppenheimer-Volkoff (TOV) equations. The input to the calculations is the EoS and the output yields the masses of the neutron star models *as a function of their radius*, for a given central energy density. To form a true test of theory accurate knowledge of both the gravitational mass and radius of the same neutron star is needed. The impossibility of separately obtaining mass and radius from the TOV equation is the main obstacle in relating observation to theory; a very difficult task which has not been satisfactorily solved. To demonstrate the present situation the four neutron stars having the most accurately measured masses and their periods of rotation are listed in Table 1.

| | | | |
|---|---|---|---|
| PSR B1913+16 | ( neutron star binary); | M = 1.4414±0.0002 $M_\odot$ | P = 59 ms |
| PSR J0737-3039 | (double pulsar (A,B)); | M = 1.249±0.001 $M_\odot$ | P = 2.77s |
| PSR J1614-2230 | (neutron star + white dwarf); | M = 1.97±0.04 $M_\odot$ | P = 3.15 ms |
| PSR J1903+0327 | neutron star + main sequence star ); | M = 1.667±0.021 $M_\odot$: | P =2.5ms |

*Table 1 Mass and rotation period of four selected neutron stars. Data taken from Refs 6 - 9.*

The solutions of the TOV equations for two very different EoS, are shown in Figure 3. The left curve is obtained with a nucleon-only EoS, derived from the realistic AV18+δ+UIX* nucleon-nucleon potential (APR) [10], the right curve corresponds to EoS from the Quark-Meson-Coupling model (QMC) [5] assuming a full baryon octet constituting the star core matter. It is clearly seen that both EoS are compatible with all four neutron star masses. However, the predicted radii are very different and lacking this information for any of these stars, a “correct” EoS cannot be selected.

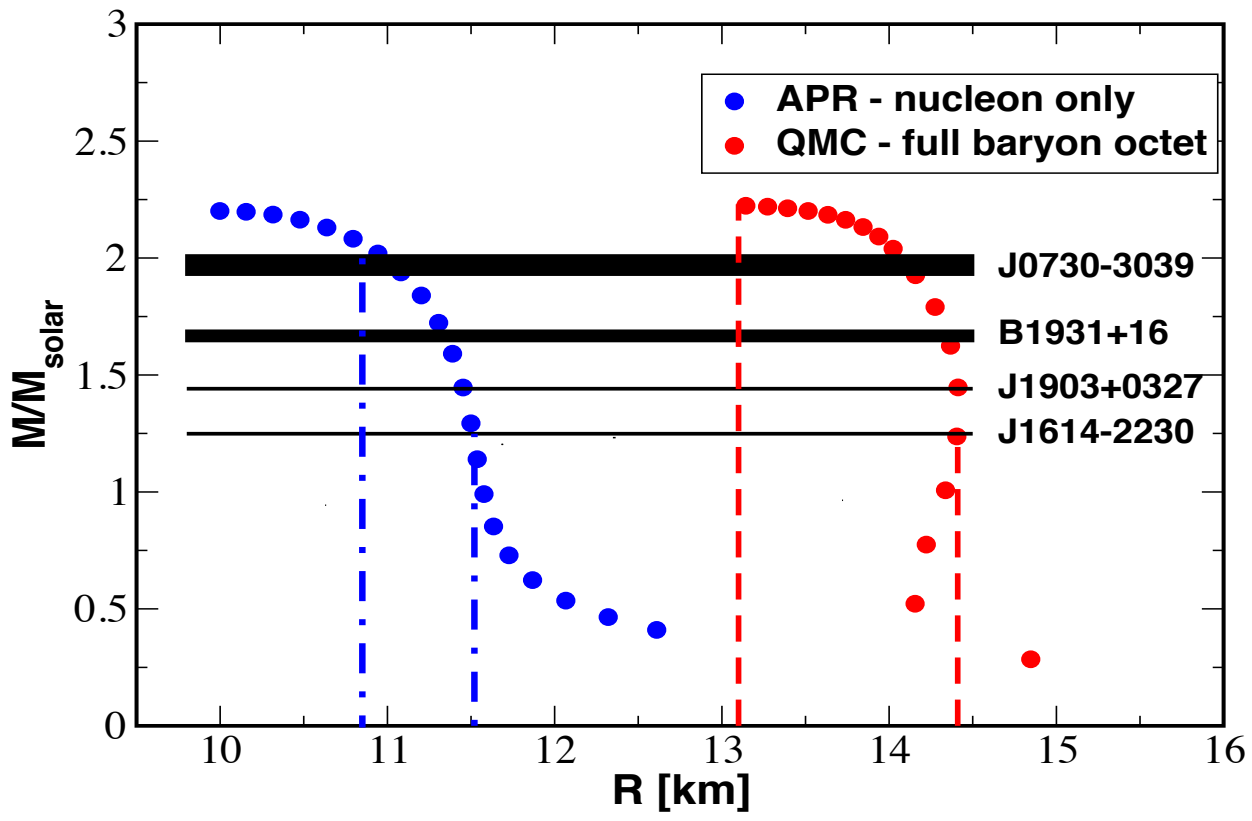

*Figure 3 M-R curves for APR and QMC EoS.*

At the time of its publication, the reportedly heaviest neutron star PSR J1614-2230 [8] claimed to rule out *almost* all then proposed hyperon or boson condensate EoS and further claimed that quark matter could support a star this massive only if the quarks were strongly interacting and therefore not 'free'. Soon after, a host of papers appeared showing that there are many EoS which support such a heavy star and contain exotic components. Moreover, compatible neutron stars with the full baryon octet of hadrons had been predicted before the observation [5,11].

Similarly, knowledge of radii of neutron stars of unknown mass also is not sufficient for selecting EoS. Radius determination from observation is difficult since it requires, among other things, information about the distance of the studied star from the Earth and angular area of the star. The best signal related to the area of the star is provided by Type I X-ray bursters and/or transient neutron stars in quiescence in globular clusters as discussed below.

**2.1.2 Simultaneous determination of gravitational mass and radii.**

Thermal radiation from cooling neutron stars, either after their birth, or after crustal heating following accretion events, is observed as X-rays. If the distance for the cooling object is known, the observed X-ray spectra can be used to determine the apparent angular emitting area, which depends on the mass and radius of the star. The most accurate data are obtained from nearby isolated cooling neutron stars or quiescent low mass X-ray binaries [12].

Type I X-ray bursts appear during a thermonuclear explosion, ignited by thermally unstable He in the accretion envelope of a neutron star from its companion. Significant variations of the radius $r_{ph}$ of the photosphere and changes of the black-body temperature during the burst can be used to set limits on mass and radius of the burster. This method was used by Ozel et al (for summary see [13]) for EXO 1745–248, 4U 1608–522, 4U 1820–30 X-ray bursters in determining mass and radius of these objects. Later the data were extended to include transient neutron stars in quiescent low mass X-ray binaries in M13 and ω Cen clusters, and X7 in 47 tuc and reanalyzed by Steiner et al. [12].

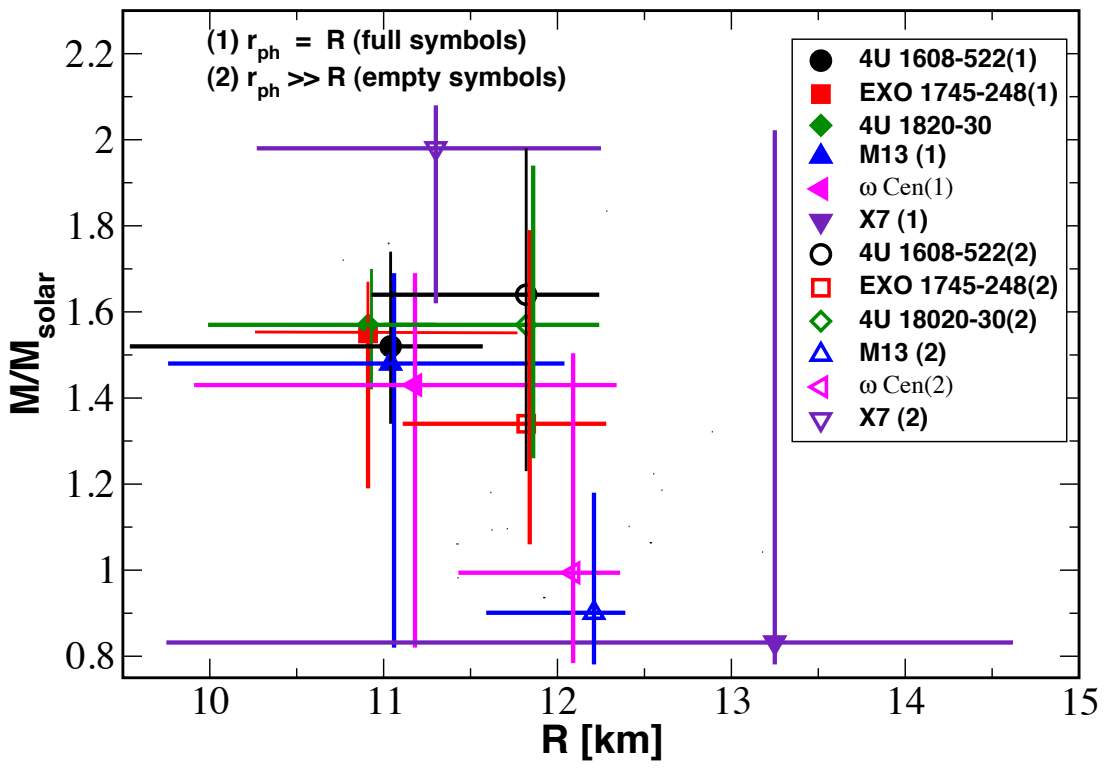

*Figure 4 Most probable values of masses and radii neutron stars as obtained in [12]. The vertical and horizontal bars represent 1σ uncertainties.*

Statistical methods were used to extract the most probable values of mass and radii of neutron stars based on the assumption that all neutron stars lie on the same M-R curve. The analysis has been performed for two cases, $r_{ph}$ = $R$ and $r_{ph}$ << $R$. The results are illustrated in Figure 4 suggesting that the radii for 1.4 -1.6 $M_\odot$ are in the region of 11-12 km. The outlying results for X7 show extreme sensitivity to the uncertainty in the photospheric radius. Fast rotation, not considered in the analysis, has been suggested as a possible cause.

Based on this analysis, Steiner et al. attempted to derive parameterized EoS with the high density part approximated by a polytrope. Such an over-simplified approach, without physical content, does not directly constrain the microscopic composition of high density matter although, for this small sample, comparison with Figure 3 would favour the APR model.

## 2.2 Core Collapse supernovae

There are three obvious signals from the CCS event: the explosion, the emission of neutrinos and the creation of a remnant in the form of either a neutron star or a black hole. The simulation of the explosion still poses a challenge to theory despite enormous efforts. The neutrino emission is, in principle, sensitive to the microscopic make-up of the supernova matter (see contribution to these Proceedings by Irina Sagert). The third signal, formation of a neutron star, is the main interest of this paper.

### 2.2.1 Proto-neutron stars.

In the first several seconds after the bounce of the collapsing core a neutrino rich proto-neutron star (PNS) begins to form. Within 0.1 - 0.5 s it shrinks from over 100 km in radius to about 15 km and keeps accreting mass from the surrounding material. At the moment of explosion the surrounding material is blown away and the PNS enters the Kelvin-Helmholz phase which lasts for several seconds. The PNS first goes through de-leptonization when excess trapped electron neutrinos diffuse from the central regions outward through the star. The temperature is still rising as the diffusing neutrinos heat the stellar core while decreasing the net lepton and proton fractions. This stage is followed by an overall cooling stage [14]. During the time period between bounce and the start of cooling the microscopic composition of the PNS is formed. It does not significantly change during the cooling period and through the life of the cold neutron star, unless the star is involved in further accretion from a companion and changes its temperature and rotational speed. Currently no established theory exists to relate the composition of the PNS to the conditions in the CCS process. Such a theory would yield interesting clues to limit the PNS make-up.

Precise data on rapid neutron star cooling have been recently reported on the isolated neutron star, with carbon atmosphere and low magnetic field, in Cassiopeia A, the remnant of the 1680 supernova [15]. The rate of cooling is significantly larger that that expected from the modified (or medium modified) URCA process. These data prompted theoretical interpretation (i) via the "minimal cooling paradigm" [16], (ii) the "nuclear medium cooling scenario" [17] and general relativistic effects [18].

Theory (i) is based on the assumption that the cooling is enhanced by neutrino emission following the onset of breaking and formation of neutron Cooper pairs in the $^3P_2$ channel in the star's core. The model assumes that the star is already in a proton superconducting state. The rapidity of the cooling is presented as firm evidence for the presence of neutron superfluidity and proton superconductivity in the core of the neutron star. Theory (ii) goes beyond the minimal cooling model and introduces medium-modified one-pion exchange in dense matter and polarization effects in the pair-breaking of neutrons and protons. The problem with these models is that the mass of the neutron star in Cas A is not certain and it is not clear whether nucleons can exists in its core. Ho and Heinke [15] estimated $M \approx 1.5$–$2.4\ M_\odot$, $R \approx 8$–$18$ km. If the mass is larger than about 1.4 $M_\odot$, it is unlikely that nucleons exist in the core and the proposition that the core makes transition to a superfluid neutron state becomes untenable. Theory (iii) approaches the rapid cooling from a very different point of view. Negreiros et al [18] developed a model of thermal evolution of rotationally deformed neutron stars in two dimensions in the framework of general relativity. They showed that the observational data can be reproduced to the same level of accuracy as is obtained by theories based on neutron superfluidity. For completeness, Yakovlev at al [19] studied cooling of Cas A without a specific assumption on microscopic structure of the star and got results compatible with observation.

The existence of several different theories explaining the rapid cooling of Cas A equally well implies that this data alone cannot distinguish between them. The mass of Cas A is required with good accuracy to confirm or exclude the neutron superfluidity theory and the period of rotation is needed to justify application of the relativistic theory of cooling to this neutron star.

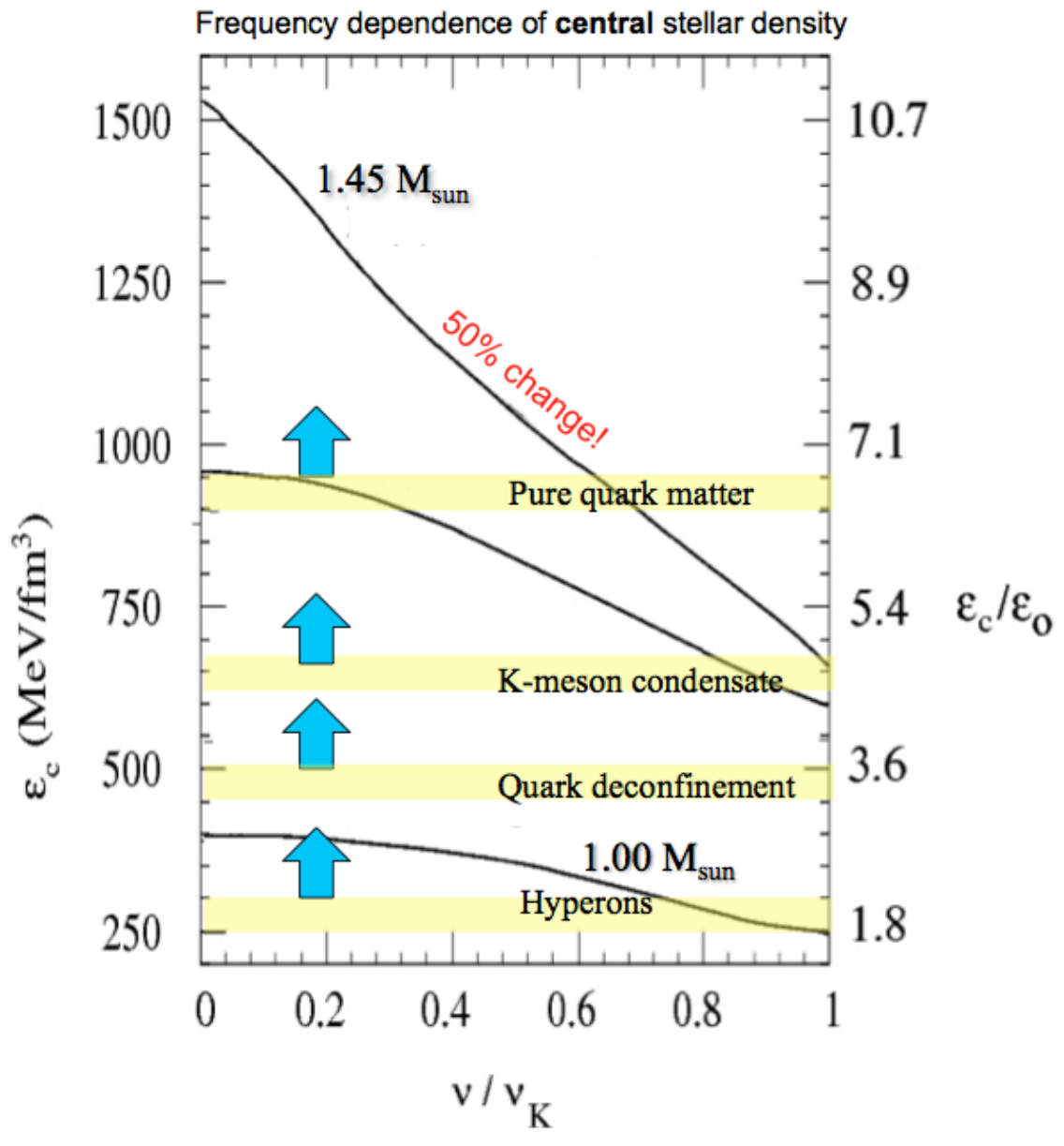

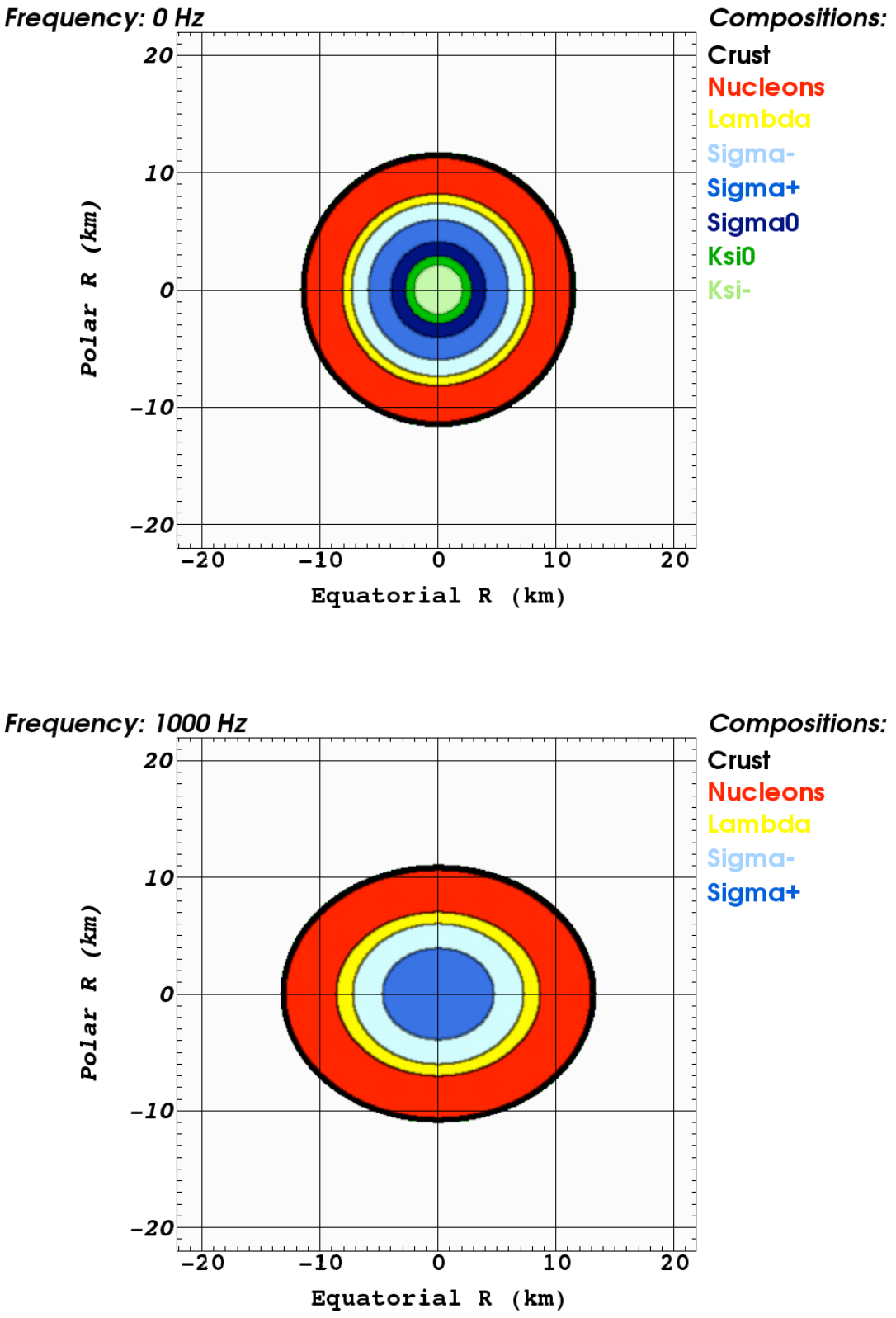

*Figure 5 (Left) Frequency dependence of central energy density of a rotating neutron star (adopted from Ref. [2]). (Right) Cartoon of composition of a the core stable neutron star and of a star rotating at 1000 MHz (thanks to Chad Pyle and Eric Lingerfelt).*

### 2.2.2 Rapidly rotating neutron stars

PNS are born rapidly rotating provided the core of the collapsing CCS star rotates, as angular momentum has to be conserved. Rotation can be induced later in the life of the star if accretion of a companion deforms the star and rotation becomes an allowed degree of freedom. In rotating stars the centrifugal force acts against gravity and thus the star may be up to 25% more massive and have about 15% larger equatorial radius than a spherical star. The most important feature of rotating stars is that the central energy density is rotation-frequency dependent. This means, for example, that the threshold energy density for appearance of hyperons will be frequency dependent, as illustrated schematically in Figure 5. The dependence has been calculated using a very soft EoS to demonstrate the effect, which would be somewhat less, but still pronounced, if a stiffer EoS were used. The baryon number was kept constant during the spin-down, as it is for isolated neutron stars. In the top right panel the schematic composition of a non-rotating star is depicted. The thickness of the layers represents the proportion of each species in the matter, starting from the outside: nucleons (red), Λ (yellow), $\Sigma^-$ (pale blue), $\Sigma^+$ (blue), $\Sigma^0$ (dark blue), $\Xi^0$ (green) and $\Xi^-$ (pale green). At 1000 Mhz (note that the frequency of the fastest known pulsar to-date is 716 Mhz) the proportions are significantly changed, with increased content of nucleons and $\Sigma^0$, $\Xi^0$ and $\Xi^-$ missing. These effects may prove significant in constraining the high density EoS. But we have to have accurate data on mass, radius and period of rotation on the same object.

### 2.3 Constraints from gravitational wave observations

Future observation of gravitational waves will open a wide range of possibilities to examine bulk properties of high-density matter. Some of the most important phenomena will include non-linear effects in spin evolution of young neutron stars [20], excitation of glitches and their post-glitch relaxation phase [21] and r-modes of accreting hyperonic stars [22].

## 3. Constraints from Heavy-Ion (A-A) Collisions[2]

Heavy-ion collisions are the only terrestial events in which hot high density matter is created. The beam energies range from ~ 35 AMeV to 5 ATeV at different facilities, MSU, Texas A&M, GSI, RHIC and LHC (existing), and FAIR and NICA (future). The signals detected include yields and energy spectra of transverse and elliptical particle flow following a collision. The data are used to fit parameters of transport models, which provide energy density and pressure in the matter, i.e. the EoS [23,24]. The much discussed question arises whether such EoS can be meaningfully used as a constraint of the EoS of high density matter in neutron stars and supernovae. We compare the main features of the form of high density matter in an A-A collision and a PNS in the panels below:

**Central A-A collision**:
Strongly beam energy dependent
Temperature: ~120 - 200 MeV
Energy density: ~ 1 - 6 GeV/fm$^3$
Baryon density < $\rho_s$
Time scale to cool-down: $10^{-(22\text{-}24)}$s
Not neutrino rich, few leptons
Strong Interaction : (S, B, L conserved)
Time scale ~$10^{-24}$ s
*Higher energies*: QGP, $q\bar{q}$ pairs,
strange baryons, pions
*Lower energies*: Inelastic NN scatterings,
N, N*, pions,
strangeness less important

QUESTION OF THERMAL AND CHEMICAL EQUILIBRIUM

**Proto-neutron star**:
Progenitor mass dependent
Temperature: < 50 MeV
Energy density: ~ 1 GeV/fm$^3$
Baryon density ~ 2-3 $\rho_s$
Time scale to cool-down: 1 -10 s
Neutrino rich matter – de-leptonized
Weak Interaction: B and L conserved
Time scale ~$10^{-10}$ s
*Higher T*: strangeness produced in
weak processes
*Lower T*: freeze-out of strangeness?
N, strange baryons,
K and π mesons and leptons

QUESTION OF THERMAL AND CHEMICAL EQUILIBRIUM

[2] Thanks to Giorgio Torrieri, Christine Nattrass and Daniel Cebra for elucidating discussions.

It is clear that conditions in the two phases of matter are very different. Although baryon number density exceeding the nuclear saturation density $\rho_s$ maybe reached in A-A collisions for a very small fraction of a second, it will rapidly decrease. The time-scale of an A-A collision implies that detection of its products can happen only after the event is over – all its phases have to be reconstructed using models. Thus any EoS extracted from transport models of A-A collisions describes low density, high temperature matter in contrast with low temperature high density matter in supernovae and PNS and, in particular, with cold and very high density matter in neutron star cores.

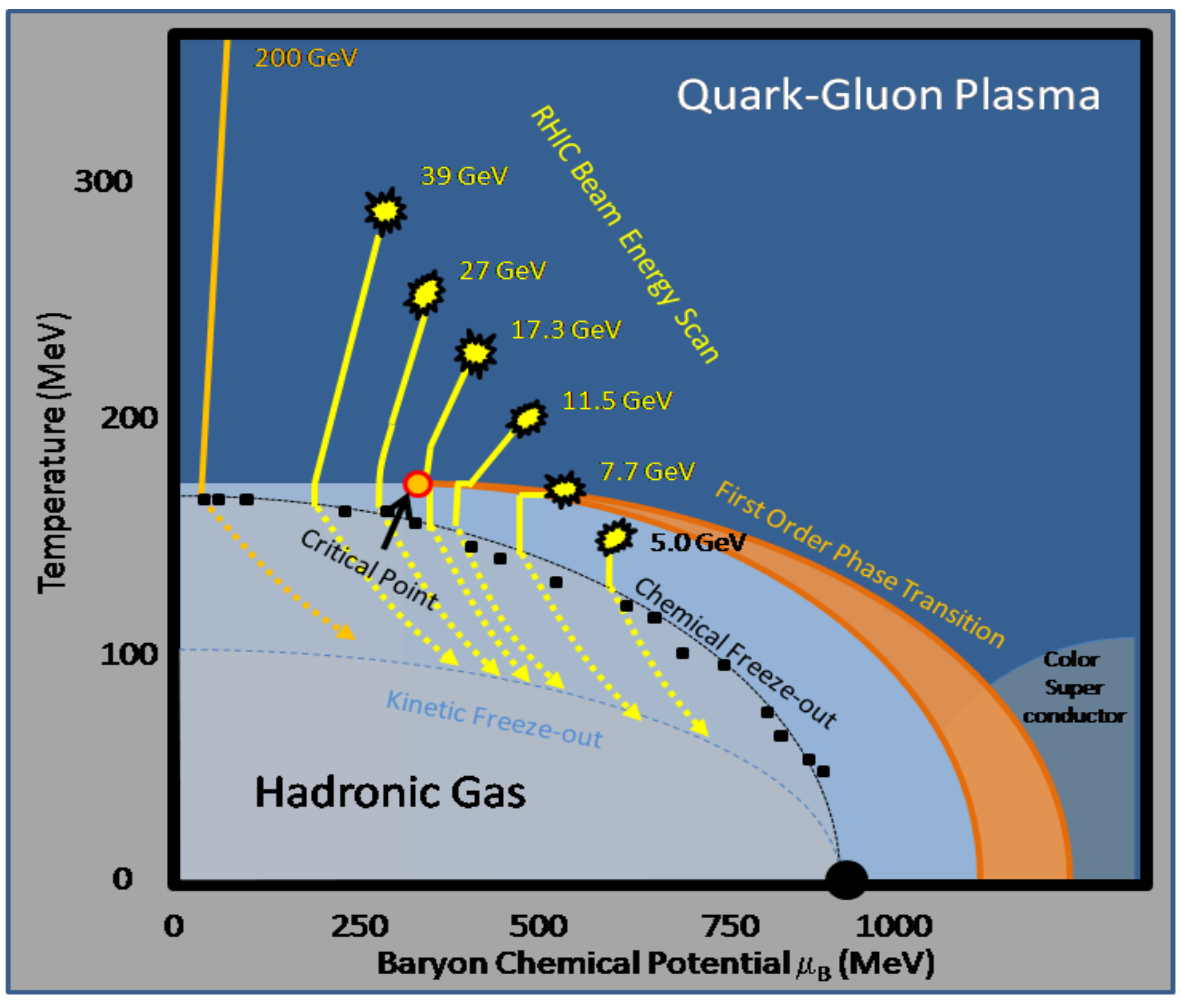

*Figure 6 Schematic RHIC beam scan progress. Taken from [25].*

One of the prime issues in heavy ion collision research is mapping of the onset of deconfinment and finding the critical point in the QCD phase diagram. Particle spectra and yield ratios, as a function of beam energy, are used to deduce T and $\mu_B$ (see Figure 6). The location of the transition between hadronic and quark phases of matter has direct implication for high density matter in astrophysical objects, i.e. the relative thresholds of appearance of hyperons and quark matter. It is plausible to speculate that, if the beam energy is low enough, conditions of matter close to that in CCS and PNS can be reached.

## 4. Summary and conclusions

The theory of high density matter draws on many areas of fundamental physics, often at their extremes. Numerous sophisticated models have been developed, but, at present, there are not enough observational/experimental data to constrain even the simplest, concerning mass, radius and rotational behaviour of neutron stars. Until more information becomes available, we are left with the following questions:

- Is there a “universal” EoS of high density matter?
- Do all neutron stars lie on the same M – R curve?
- Does tmatter in A – A collisions have the same EoS as PNS and neutron stars?
- What is the nature of the hadron – quark matter transition in hot dense matter?

**Q**: **Horst Stoecker**: At ~10 Gev per nucleon heavy ion energies, baryon densities can become very high and parts of this dense matter can be rather cool and strange, so at the FAIR-facility we expect to get high statistics data of cosmic matter in the laboratory.
**A**: This would be extremely useful. The question still remains about the time-scale of the collision and the predominance of the strong interaction during the heavy ion collision.
**Q**: **Filippo Galeazzi**: From heavy ion collision (HIC) experiments it is not possible to uniquely define a compressibility parameter for different densities. Why do all EoS available on the market use a constant value of this parameter?

**A**: This is a very good question. To my knowledge nobody considered the density dependence of the incompressibility coefficient K. The reason is that K is defined for symmetric nuclear matter at saturation density. All HIC simulations treat K as a free parameter of their fitting procedure. I do not believe this helps.

**Q**: **David Kahl**: In regard to experimental efforts – in particular heavy ion collisions – it seems to be a system very far from equilibrium and I wonder how well temperature can be defined in that circumstance. What are your thoughts on this?

**A**. A very good comment. We understand temperature as a property of a system in equilibrium. Thermal equilibrium is a very much-discussed issue by the HIC community and the definitive answer is not known.